\documentclass[aps,prb,preprint,superscriptaddress]{revtex4-1}
\usepackage{graphicx}
\usepackage{dcolumn}
\usepackage{bm}

\begin{document}


\title{Left-handed Band in an Electromagnetic Metamaterial Induced by Sub-wavelength Multiple Scattering}
\author{Simon Yves}
\affiliation{Institut Langevin, ESPCI Paris, PSL Research University, CNRS UMR 7587,  1 rue Jussieu, 75005 Paris, France}
\author{Thomas Berthelot}
\affiliation{CEA Saclay, IRAMIS, NIMBE, LICSEN, UMR 3685, F-91191, Gif sur Yvette, France}
\affiliation{KELENN Technology, Antony, France}
\author{Mathias Fink}
\affiliation{Institut Langevin, ESPCI Paris, PSL Research University, CNRS UMR 7587,  1 rue Jussieu, 75005 Paris, France}
\author{Geoffroy Lerosey}
\affiliation{Greenerwave, ESPCI Paris Incubator PC’up, 6 rue Jean Calvin, 75005 Paris, France}
\author{Fabrice Lemoult}
\affiliation{Institut Langevin, ESPCI Paris, PSL Research University, CNRS UMR 7587,  1 rue Jussieu, 75005 Paris, France}
\email[]{fabrice.lemoult@espci.fr}

\date{\today}

\begin{abstract}
Due to the deep sub-wavelength unit cell in metamaterials, the quasi-static approximation is usually employed to describe the propagation. By making pairs of resonators, we highlight that multiple scattering also occurs at this scale and results in the existence of a dipolar resonance, which leads to a negative index of refraction when we consider several resonators. We experimentally verify the possibility of obtaining a negative index of refraction in periodic metamaterials in two different ways, and eventually demonstrates a superlensing effect in both cases.
\end{abstract}

\maketitle

The wave-matter interaction is the basis for many applications relying on the wave-like nature of the system. Indeed, from the waves scattered by an object we can extract a specific property: for example the X-ray diffraction off the Bragg planes of a crystal allows to recover its crystalline structure~\cite{warren1969}, the infra-red spectroscopy gives information about the chemical composition of a flask through molecular resonance excitations~\cite{Stuart2005}, or even magnetic resonance imaging builds medical images from mapping relaxation times of atoms~\cite{Lauterbur1973}. All of these understandings of the wave-matter interaction have recently lead to the advent of a new class of artifical media: the {\it metamaterials}. In this case, the strategy is slightly different; we engineer artificial objects that interact with waves in order to induce a new macroscopic property~\cite{Engheta2006,Cai2010,Capolino2009}. The motor of many advances in the field has been the quest for left-handed media exhibiting negative refraction~\cite{Veselago1968,Smith2004}, because of the inital proposal that a slab made of such a medium should behave as a perfect lens~\cite{Pendry2000}. Nevertheless, this does not limit to this example, and many other ideas emerged as for example proposals of cloaking~\cite{Pendry2006,Leonhardt2006,Schurig2006}, high-effective index~\cite{Choi2011}, or epsilon-near-zero media~\cite{Engheta2013,Alu2007}.

The double negativity of a system has often been obtained by mixing two different kinds of media, one bringing the negative effective permittivity~\cite{Pendry1996}, and the other one being responsible of the negative permeability~\cite{Pendry1999}. This approach neglects all of the possible interactions at the unit cell level between the two kinds of media. Recently, in acoustics, it has been demonstrated that a single negative metamaterial based on a resonant unit cell can be turned into a double negative one if the resonators are grouped by pairs~\cite{Kaina2015,Lanoy2017}. Typically, multiple scattering which occurs between the two-resonators creates a dipolar mode that spectrally overlaps with the initial monopolar one resulting in a double negative property. In a periodic medium, a very convenient way to build physically a bigger super-cell that contains two resonators consists in either detuning one resonator out of two, thus ending on a bi-disperse medium, or by slightly moving the position of this resonator and ending on a bi-periodic medium. In this article, We experimentally explore these two different strategies in the microwave domain.

\begin{figure}[bt]
\includegraphics{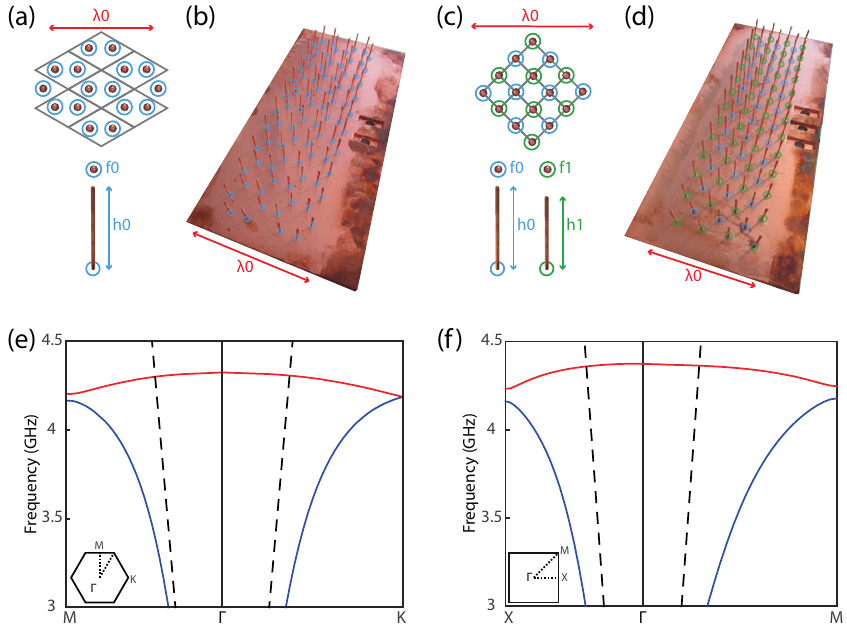}
	\caption{\label{Fig1} (a) The bi-periodic metamaterial consisting on a honey-comb lattice of identical metallic wires on a ground plane, and (b) the corresponding experimental sample. (c) The bi-disperse one based on a square lattice containing two wires of slightly different lengths, and (d) its experimental realization. (e) and (f) the numerical dispersion relations along the principal directions for respectively the bi-periodic and the bi-disperse metamaterials (in red the negative branch).}
\end{figure}

In our study, the resonant unit cell consists in a simple quarter-wavelength metallic rod on a ground plane. Seen from the top each of these resonators is very small compared to the wavelength and their resonant nature comes from their length. Here, we choose a rod diameter of 1~mm and a height of $h_0=$16~mm, therefore resonating around 4.68~GHz. As a consequence, we can pack many of such elementary bricks within a wavelength and we therefore directly fall within the context of locally resonant metamaterials. It is now well known that the finite-length wire medium can be described with an effective permittivity which presents a Lorentzian profile near each resonance~\cite{Simovski2012}. Basically with a metamaterial approach we would assume that all the resonators see the same incident electric field and their overall response is averaged in order to give a macroscopic effective property that only depends on the number of resonators per wavelength. With this description, the spatial organization of the wires does not influence the macroscopic behavior, but only their density does. 

Now, we want to induce a dipolar resonance by building a metamaterial that is still based on pairs of the previous resonant unit cell. We can do it in the two different aforementioned ways: the bi-periodic medium and the bi-dispersed one. In two dimensions, the most isotropic bi-periodic medium consists in a honey-comb arrangement of resonators (Fig.\ref {Fig1}.a), and our choice of bi-dispersed sample consists of a square lattice with two detuned resonators per unit cell (\ref {Fig1}.c, the shorter wires have now a height $h_1=$15.7~mm).  For both cases, we build a sample by metallizing a 3D-printed ABS-based plastic medium with the same procedure as in the ref.~\cite{Kaina2017}. In each case, nearest neighboring rods are separated by a distance of 10~mm. The two fabricated samples are shown in figure~\ref{Fig1}.b and d and have the shape of a rectangle paving an area of 160~mm~x~60~mm. 

For each lattice, the dispersion relation is computed numerically using Comsol Multiphysics along the main directions of the crystals by simulating a single unit cell with Bloch periodic boundary conditions (Fig.~\ref{Fig1}.e and f). As expected, both of them clearly reveal the existence of a band with a negative gradient within the first Brillouin zone (red line). The negativeness of this slope is a direct consequence of the existence of a dipolar resonance overlapping with a monopolar one, {\it ie.} the two eigenmodes of the unit cell based on two resonators. This first numerical result confirms that, albeit the deep sub-wavelength step between the two resonators, a dipolar mode that exhibits a change of sign between the them exists: the usual description of such a metamaterial that solely takes into account a monopolar one is obviously not sufficient.   

\begin{figure}[bt]
	\includegraphics{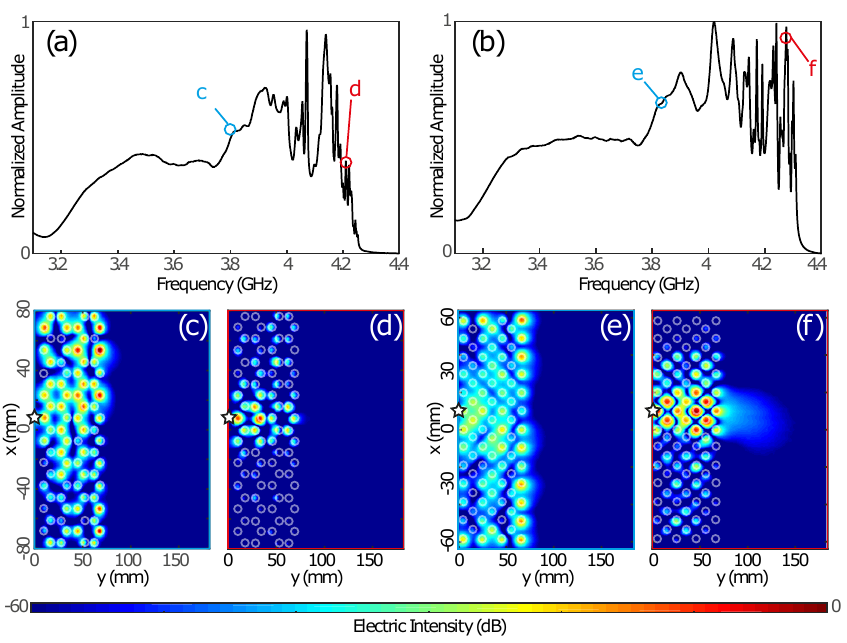}
	\caption{ \label{Fig2} {\bf Point source excitation.} (a) (respectively (b)) A plot of a typical measured spectrum on top of wire within the metamaterial when excited by a monopole antenna near one interface for the bi-periodic (respectively bi-disperse) sample. (c-f) Intensity maps on top of the medium for one frequency within the first band (c and e) and one frequency in the second one (d and f) for both samples.}
\end{figure}

In order to verify that the presence of this band traduces a negative refraction phenomenon, we perform a point source excitation at one interface of the slab samples. In both cases, we measure the transmission, using a vector network analyser, between this source and a near-field probe placed in the middle of the sample. This probe is mounted on a motorized 2D-stage and will be used later to map the field on top of the slab. Typical frequency spectra measured for both samples are represented in figure~\ref{Fig2}.a and b. 

Both of them reveal the presence of resonance peaks, spanning frequencies from 3.5~GHz to 4.3~GHz, falling within the previously calculated propagating bands. The linewidth of the resonances seems to narrow as approaching the intrinsic resonance $f_0$ of the wires as in any finite-sized locally resonant metamaterial~\cite{Lemoult2010,Lemoult2013,Lemoult2011a}, but we cannot yet see if the spatial organization of the rods has created any specific effect.

A first step to evidence the negative refraction comes from our two dimensional scan. Indeed, we display in figure~\ref{Fig2}.c-f the intensity maps corresponding to different frequencies. They can be classified in two very distinct groups. In the lower part of the resonance peaks (c and e) the intensity is distributed over the entire slab, while for the peaks near the band gap (d and f) the intensity seems to be confined along the horizontal axis facing the source position. A ray tracing with a negative refraction occurring at each interface is tempting to explain this effect, but canalization~\cite{Belov2005} or anisotropic efficiency of the bandgap could also cause such an intensity distribution. In any case, the intensity maps lack the phase information that is required to clearly identify if those metamaterials could be described with a negative index of refraction.

\begin{figure}[bt]
\includegraphics{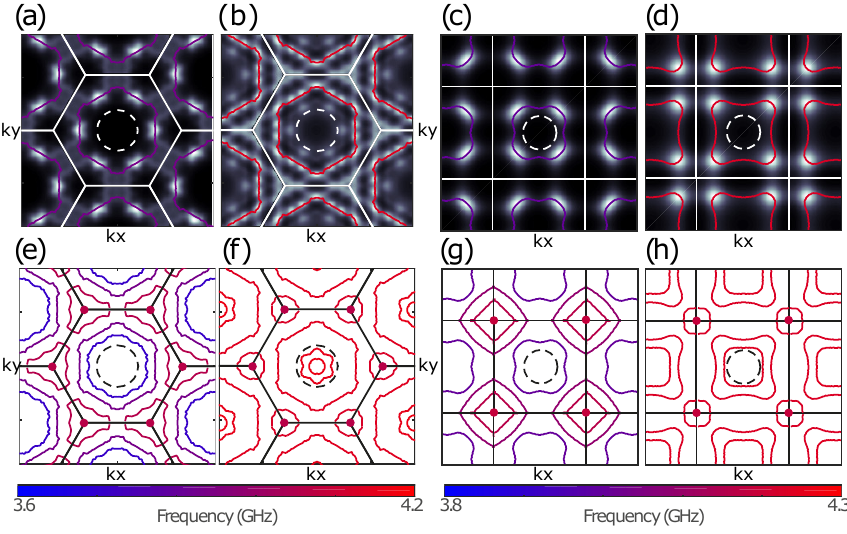}
	\caption{\label{Fig3} {\bf Experimental isofrequency contours.} (a-d) Representation in the reciprocal space of the field maps of Fig.~\ref{Fig2} (see text for details). Superimposed in white are the first Brillouin zone and its periodic replica (an hexagon for the bi-periodic crystal and a square for the bi-disperse one), and in color the extracted isofrequency contour for this frequency. (e-g) The isofrequency contours obtained when processing all the measured maps. (f) and (h) corresponds to the two desired negative branches.}
\end{figure}

A convenient representation to extract the effective properties of a medium at a single frequency is the reciprocal space. We therefore carry out two dimensional Fourier transforms of the field maps and show the intensity distribution as a function of the wavevector $\mathbf{k}$. Due to the slab geometry this leads to anisotropic intensity distribution. We therefore perform some symmetry operations on the resulting map. For the case of the honey-comb lattice they consist in applying a C6 rotational symmetry and 2 mirrors (vertical and horizontal). For the square lattice made of two different resonators these operations are a C4 rotational symmetry combined with the two mirrors. In both cases we also take into account the translational symmetry with respect to the first Brillouin zone. Once these operations are done on the maps initially represented in intensity in the real space (figure~\ref{Fig2}.c-f), we end up on the intensity maps represented in the reciprocal space of figure~\ref{Fig3}.a-f. For convenience we superimpose in white the first Brillouin zone as well as its periodic replica. We also superimpose a line that corresponds to an automatic detection of the intensity maximum in all propagating directions. For example, in figure~\ref{Fig3}.a it results in a rounded contour which still shows a C6 symmetry but can fairly be approximated by a circle. The radius of this circle is bigger than the free space circle (white dotted line) meaning that the dispersion relation at this frequency can be approximated by an effective index of refraction higher than unity in norm. For figure~\ref{Fig3}.b the isofrequency contour looks more like an hexagon with rounded edges than a circle meaning that some anisotropy effects exist, but again this contour is outside of the freespace one (dashed line). Similar observations can be made for the sub-panels c and d except that in this case the reminiscence of the C4 symmetry of the crystal persists.  At any rate, even if modeling the propagation in the metamaterial in terms of a unique scalar, {\it ie.} the effective index refraction, is probably  too restrictive due to the anisotropy, this figure does not permit to conclude regarding the sign of this index of refraction.

We applied the same procedure for several frequencies and plot all of the obtained isofrequency contours. We separate the results in two different graphs for each type of crystal since they can be associated to the two desired bands with very different properties. For the honey-comb lattice case, the lowest frequency (blue line in figure~\ref{Fig3}.e) corresponds to a circle slightly bigger than the freespace one. Increasing the frequency (meaning changing the color from blue to violet) we notice a rapid expansion of the circle's radius as expected from the usual picture of the polariton. Another frequency increment leads to an anistropic behavior and the propagation becomes forbidden in the $\Gamma$M directions while authorized in the $\Gamma$K directions. This is an unambiguous evidence of the crystalline nature of the propagation within such a metamaterial, albeit its deep sub-wavelength spatial scale. Moreover, even if this is out of the scope of this article, our measurements reveal the existence of a Dirac cone at the K-point~\cite{Geim2010}, which is far away from the free space light cone. Going to even higher frequencies, we move to the second propagation band and to the sub-panel (f) of the figure. This corresponds to the desired effect: after few frequencies still exhibiting some anisotropic effects the isofrequency contours are circles centered on $\Gamma$ with a radius that decreases with frequency. The circle-shape indicates that the propagation can be well described with a single scalar (the index of refraction) while the decrease in the radius with frequency indicates the negativeness of the latter. Very similar conclusions can be made on the isofrequency contours extracted from the measurement on the square lattice of rods (panel g and h) except that the anisotropy is more pronounced since the shape of the contours is more similar to rounded squares.

\begin{figure}[bt]
\includegraphics{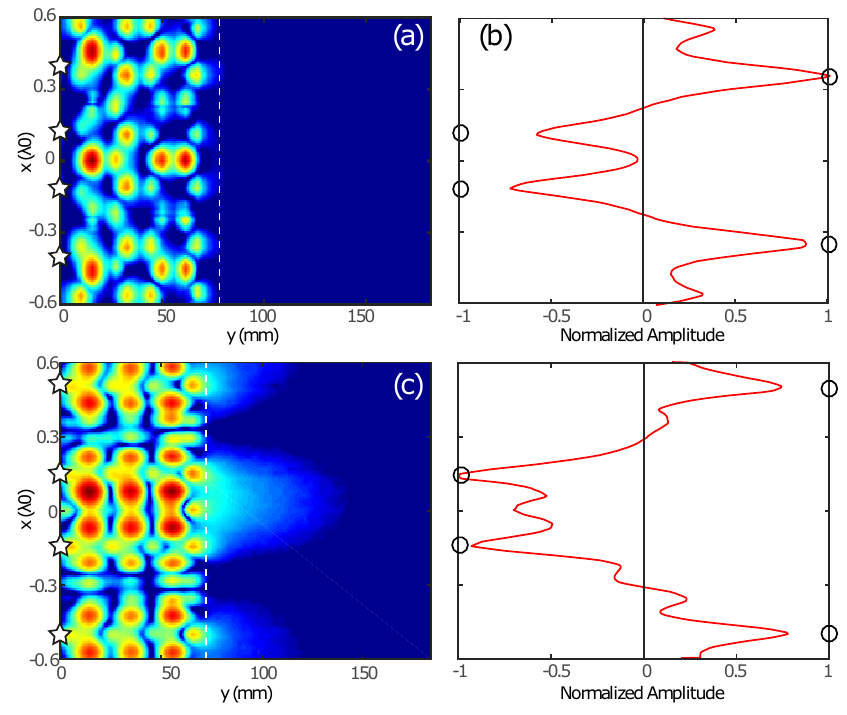}
	\caption{\label{Fig4} {\bf Sub-wavelength imaging}. For both crystals the source is now a simultaneous emission of 4 point-antennas separated by a distance of $\lambda_0/3$. We extracted the field oat the output interface shown in white dashed-line on the measured intensity maps for the bi-periodic (a) and the bi-disperse (c) samples. In both cases the reconstructed object (b) and (d) exhibits superlensing since the initial profile (circles) is reconstructed below the diffraction limit.}	
\end{figure}

Let us now move to the last experiment which results from these observations. Since the seminal work of J. Pendry~\cite{Pendry2000}, the quest for the negative index of refraction has been motivated by the superlensing effect. Namely, he stated that a flat layer of a negative index material not only works as a lens but also allows to beat the diffraction limit. This is typically what we have tried to perform for both our slabs similarly to underwater acoustic experiments~\cite{robillard2011,sukhovich2009}. The experiment consists now in emitting a sub-wavelength-scaled pattern on one face of the slab and we map the field on the other interface. Typically, four point sources linearly placed with a separation distance around $\lambda_0/3$ emit simultaneously, at the operating frequency of 4.2~GHz for the honey-comb sample, and of 4.27~GHz for the square bi-disperse one. For the two experiments this corresponds to a frequency for which we have obtained a circle-like isofrequency contour. The real part of the measured electric field at the output interface (Fig.~\ref{Fig4} b and d) strongly demonstrates that the flat lens allows to image our initial object (black circle). Even the nearest neighbors separated by a distance of $\lambda_0/3$ are resolved. This effect here is actually a consequence of the high absolute value of the index of refraction within the slabs as in solid immersion lenses~\cite{Mansfield1990} for example. The superlensing effect in those two examples therefore comes from the conversion of the evanescent waves at the input into propagating waves within the slab, and they convert again to evanescent ones at the output thus recovering small details of the initial source. There is actually no evanescent amplification of the field within the metamaterial despite the description in terms of a negative index.

As a summary, in this letter we have experimentally validated in microwaves the new strategy proposed in~\cite{Kaina2015} to build a negative index metamaterial starting from a medium with solely one negative effective property (here the permittivity). This is simply based upon the idea that a bigger unit cell made of a pair of resonators should support a dipolar resonance. Building such a meta-molecule is performed either by creating a bi-periodic medium (by moving the position of one resonator out of two) or a bi-periodic one (by detuning the same resonator), thus resulting in a periodic metamaterial whose unit cell contains two resonators. This experimental study unambiguously demonstrates that in both cases the multiple scattering induced dipolar resonance results in negative refraction. Thanks to a complete mapping of the field within our samples we experimentally extracted the full band diagram and strongly confirmed that a description in terms of a negative effective index of refraction is relevant for a given frequency range. Eventually, we observed a superlensing effect in both our samples for frequencies where the index of refraction is negative but with an absolute value higher than one.

All of these results clearly show that the spatial organization of the resonators within a metamaterial plays a crucial role, albeit the small separation distance between them with respect to the free space wavelength. We cannot neglect anymore the correlation in positions between the resonators~\cite{Lanoy2017} and this effect is notoriously intensified in periodic media. There is no doubt that more and more complex unit cells~\cite{Yves2017} lead to new exotic macroscopic effective properties in the here-opened context of crystalline metamaterials.

\bibliography{biblio}

\begin{thebibliography}{30}%
\makeatletter
\providecommand \@ifxundefined [1]{%
 \@ifx{#1\undefined}
}%
\providecommand \@ifnum [1]{%
 \ifnum #1\expandafter \@firstoftwo
 \else \expandafter \@secondoftwo
 \fi
}%
\providecommand \@ifx [1]{%
 \ifx #1\expandafter \@firstoftwo
 \else \expandafter \@secondoftwo
 \fi
}%
\providecommand \natexlab [1]{#1}%
\providecommand \enquote  [1]{``#1''}%
\providecommand \bibnamefont  [1]{#1}%
\providecommand \bibfnamefont [1]{#1}%
\providecommand \citenamefont [1]{#1}%
\providecommand \href@noop [0]{\@secondoftwo}%
\providecommand \href [0]{\begingroup \@sanitize@url \@href}%
\providecommand \@href[1]{\@@startlink{#1}\@@href}%
\providecommand \@@href[1]{\endgroup#1\@@endlink}%
\providecommand \@sanitize@url [0]{\catcode `\\12\catcode `\$12\catcode
  `\&12\catcode `\#12\catcode `\^12\catcode `\_12\catcode `\%12\relax}%
\providecommand \@@startlink[1]{}%
\providecommand \@@endlink[0]{}%
\providecommand \url  [0]{\begingroup\@sanitize@url \@url }%
\providecommand \@url [1]{\endgroup\@href {#1}{\urlprefix }}%
\providecommand \urlprefix  [0]{URL }%
\providecommand \Eprint [0]{\href }%
\providecommand \doibase [0]{http://dx.doi.org/}%
\providecommand \selectlanguage [0]{\@gobble}%
\providecommand \bibinfo  [0]{\@secondoftwo}%
\providecommand \bibfield  [0]{\@secondoftwo}%
\providecommand \translation [1]{[#1]}%
\providecommand \BibitemOpen [0]{}%
\providecommand \bibitemStop [0]{}%
\providecommand \bibitemNoStop [0]{.\EOS\space}%
\providecommand \EOS [0]{\spacefactor3000\relax}%
\providecommand \BibitemShut  [1]{\csname bibitem#1\endcsname}%
\let\auto@bib@innerbib\@empty
\bibitem [{\citenamefont {Warren}(1969)}]{warren1969}%
  \BibitemOpen
  \bibfield  {author} {\bibinfo {author} {\bibfnamefont {B.~E.}\ \bibnamefont
  {Warren}},\ }\href@noop {} {\emph {\bibinfo {title} {X-ray Diffraction}}}\
  (\bibinfo  {publisher} {Courier Corporation},\ \bibinfo {year}
  {1969})\BibitemShut {NoStop}%
\bibitem [{\citenamefont {Stuart}(2005)}]{Stuart2005}%
  \BibitemOpen
  \bibfield  {author} {\bibinfo {author} {\bibfnamefont {B.}~\bibnamefont
  {Stuart}},\ }\href@noop {} {\emph {\bibinfo {title} {Infrared
  spectroscopy}}}\ (\bibinfo  {publisher} {Wiley Online Library},\ \bibinfo
  {year} {2005})\BibitemShut {NoStop}%
\bibitem [{\citenamefont {Lauterbur}\ \emph {et~al.}(1973)\citenamefont
  {Lauterbur} \emph {et~al.}}]{Lauterbur1973}%
  \BibitemOpen
  \bibfield  {author} {\bibinfo {author} {\bibfnamefont {P.~C.}\ \bibnamefont
  {Lauterbur}} \emph {et~al.},\ }\href@noop {} {\bibfield  {journal} {\bibinfo
  {journal} {Nature}\ }\textbf {\bibinfo {volume} {242}},\ \bibinfo {pages}
  {190} (\bibinfo {year} {1973})}\BibitemShut {NoStop}%
\bibitem [{\citenamefont {Engheta}\ and\ \citenamefont
  {Ziolkowski}(2006)}]{Engheta2006}%
  \BibitemOpen
  \bibfield  {author} {\bibinfo {author} {\bibfnamefont {N.}~\bibnamefont
  {Engheta}}\ and\ \bibinfo {author} {\bibfnamefont {R.~W.}\ \bibnamefont
  {Ziolkowski}},\ }\href@noop {} {\emph {\bibinfo {title} {Metamaterials:
  physics and engineering explorations}}}\ (\bibinfo  {publisher} {John Wiley
  \& Sons},\ \bibinfo {year} {2006})\BibitemShut {NoStop}%
\bibitem [{\citenamefont {Cai}\ and\ \citenamefont {Shalaev}(2010)}]{Cai2010}%
  \BibitemOpen
  \bibfield  {author} {\bibinfo {author} {\bibfnamefont {W.}~\bibnamefont
  {Cai}}\ and\ \bibinfo {author} {\bibfnamefont {V.~M.}\ \bibnamefont
  {Shalaev}},\ }\href@noop {} {\emph {\bibinfo {title} {Optical
  metamaterials}}},\ Vol.~\bibinfo {volume} {10}\ (\bibinfo  {publisher}
  {Springer},\ \bibinfo {year} {2010})\BibitemShut {NoStop}%
\bibitem [{\citenamefont {Capolino}(2009)}]{Capolino2009}%
  \BibitemOpen
  \bibfield  {author} {\bibinfo {author} {\bibfnamefont {F.}~\bibnamefont
  {Capolino}},\ }\href@noop {} {\emph {\bibinfo {title} {Theory and phenomena
  of metamaterials}}}\ (\bibinfo  {publisher} {CRC press},\ \bibinfo {year}
  {2009})\BibitemShut {NoStop}%
\bibitem [{\citenamefont {Veselago}(1968)}]{Veselago1968}%
  \BibitemOpen
  \bibfield  {author} {\bibinfo {author} {\bibfnamefont {V.~G.}\ \bibnamefont
  {Veselago}},\ }\href@noop {} {\bibfield  {journal} {\bibinfo  {journal}
  {Soviet physics uspekhi}\ }\textbf {\bibinfo {volume} {10}},\ \bibinfo
  {pages} {509} (\bibinfo {year} {1968})}\BibitemShut {NoStop}%
\bibitem [{\citenamefont {Smith}\ \emph {et~al.}(2004)\citenamefont {Smith},
  \citenamefont {Pendry},\ and\ \citenamefont {Wiltshire}}]{Smith2004}%
  \BibitemOpen
  \bibfield  {author} {\bibinfo {author} {\bibfnamefont {D.~R.}\ \bibnamefont
  {Smith}}, \bibinfo {author} {\bibfnamefont {J.~B.}\ \bibnamefont {Pendry}}, \
  and\ \bibinfo {author} {\bibfnamefont {M.~C.~K.}\ \bibnamefont {Wiltshire}},\
  }\href {\doibase 10.1126/science.1096796} {\bibfield  {journal} {\bibinfo
  {journal} {Science}\ }\textbf {\bibinfo {volume} {305}},\ \bibinfo {pages}
  {788} (\bibinfo {year} {2004})}\BibitemShut {NoStop}%
\bibitem [{\citenamefont {Pendry}(2000)}]{Pendry2000}%
  \BibitemOpen
  \bibfield  {author} {\bibinfo {author} {\bibfnamefont {J.~B.}\ \bibnamefont
  {Pendry}},\ }\href {\doibase 10.1103/PhysRevLett.85.3966} {\bibfield
  {journal} {\bibinfo  {journal} {Phys. Rev. Lett.}\ }\textbf {\bibinfo
  {volume} {85}},\ \bibinfo {pages} {3966} (\bibinfo {year}
  {2000})}\BibitemShut {NoStop}%
\bibitem [{\citenamefont {Pendry}\ \emph {et~al.}(2006)\citenamefont {Pendry},
  \citenamefont {Schurig},\ and\ \citenamefont {Smith}}]{Pendry2006}%
  \BibitemOpen
  \bibfield  {author} {\bibinfo {author} {\bibfnamefont {J.~B.}\ \bibnamefont
  {Pendry}}, \bibinfo {author} {\bibfnamefont {D.}~\bibnamefont {Schurig}}, \
  and\ \bibinfo {author} {\bibfnamefont {D.~R.}\ \bibnamefont {Smith}},\
  }\href@noop {} {\bibfield  {journal} {\bibinfo  {journal} {Science}\ }\textbf
  {\bibinfo {volume} {312}},\ \bibinfo {pages} {1780} (\bibinfo {year}
  {2006})}\BibitemShut {NoStop}%
\bibitem [{\citenamefont {Leonhardt}(2006)}]{Leonhardt2006}%
  \BibitemOpen
  \bibfield  {author} {\bibinfo {author} {\bibfnamefont {U.}~\bibnamefont
  {Leonhardt}},\ }\href@noop {} {\bibfield  {journal} {\bibinfo  {journal}
  {Science}\ }\textbf {\bibinfo {volume} {312}},\ \bibinfo {pages} {1777}
  (\bibinfo {year} {2006})}\BibitemShut {NoStop}%
\bibitem [{\citenamefont {Schurig}\ \emph {et~al.}(2006)\citenamefont
  {Schurig}, \citenamefont {Mock}, \citenamefont {Justice}, \citenamefont
  {Cummer}, \citenamefont {Pendry}, \citenamefont {Starr},\ and\ \citenamefont
  {Smith}}]{Schurig2006}%
  \BibitemOpen
  \bibfield  {author} {\bibinfo {author} {\bibfnamefont {D.}~\bibnamefont
  {Schurig}}, \bibinfo {author} {\bibfnamefont {J.~J.}\ \bibnamefont {Mock}},
  \bibinfo {author} {\bibfnamefont {B.~J.}\ \bibnamefont {Justice}}, \bibinfo
  {author} {\bibfnamefont {S.~A.}\ \bibnamefont {Cummer}}, \bibinfo {author}
  {\bibfnamefont {J.~B.}\ \bibnamefont {Pendry}}, \bibinfo {author}
  {\bibfnamefont {A.~F.}\ \bibnamefont {Starr}}, \ and\ \bibinfo {author}
  {\bibfnamefont {D.~R.}\ \bibnamefont {Smith}},\ }\href {\doibase
  10.1126/science.1133628} {\bibfield  {journal} {\bibinfo  {journal}
  {Science}\ }\textbf {\bibinfo {volume} {314}},\ \bibinfo {pages} {977}
  (\bibinfo {year} {2006})}\BibitemShut {NoStop}%
\bibitem [{\citenamefont {Choi}\ \emph {et~al.}(2011)\citenamefont {Choi},
  \citenamefont {Lee}, \citenamefont {Kim}, \citenamefont {Kang}, \citenamefont
  {Shin}, \citenamefont {Kwak}, \citenamefont {Kang}, \citenamefont {Lee},
  \citenamefont {Park},\ and\ \citenamefont {Min}}]{Choi2011}%
  \BibitemOpen
  \bibfield  {author} {\bibinfo {author} {\bibfnamefont {M.}~\bibnamefont
  {Choi}}, \bibinfo {author} {\bibfnamefont {S.~H.}\ \bibnamefont {Lee}},
  \bibinfo {author} {\bibfnamefont {Y.}~\bibnamefont {Kim}}, \bibinfo {author}
  {\bibfnamefont {S.~B.}\ \bibnamefont {Kang}}, \bibinfo {author}
  {\bibfnamefont {J.}~\bibnamefont {Shin}}, \bibinfo {author} {\bibfnamefont
  {M.~H.}\ \bibnamefont {Kwak}}, \bibinfo {author} {\bibfnamefont {K.-Y.}\
  \bibnamefont {Kang}}, \bibinfo {author} {\bibfnamefont {Y.-H.}\ \bibnamefont
  {Lee}}, \bibinfo {author} {\bibfnamefont {N.}~\bibnamefont {Park}}, \ and\
  \bibinfo {author} {\bibfnamefont {B.}~\bibnamefont {Min}},\ }\href@noop {}
  {\bibfield  {journal} {\bibinfo  {journal} {Nature}\ }\textbf {\bibinfo
  {volume} {470}},\ \bibinfo {pages} {369} (\bibinfo {year}
  {2011})}\BibitemShut {NoStop}%
\bibitem [{\citenamefont {Engheta}(2013)}]{Engheta2013}%
  \BibitemOpen
  \bibfield  {author} {\bibinfo {author} {\bibfnamefont {N.}~\bibnamefont
  {Engheta}},\ }\href {\doibase 10.1126/science.1235589} {\bibfield  {journal}
  {\bibinfo  {journal} {Science}\ }\textbf {\bibinfo {volume} {340}},\ \bibinfo
  {pages} {286} (\bibinfo {year} {2013})}\BibitemShut {NoStop}%
\bibitem [{\citenamefont {Alu}\ \emph {et~al.}(2007)\citenamefont {Alu},
  \citenamefont {Silveirinha}, \citenamefont {Salandrino},\ and\ \citenamefont
  {Engheta}}]{Alu2007}%
  \BibitemOpen
  \bibfield  {author} {\bibinfo {author} {\bibfnamefont {A.}~\bibnamefont
  {Alu}}, \bibinfo {author} {\bibfnamefont {M.~G.}\ \bibnamefont
  {Silveirinha}}, \bibinfo {author} {\bibfnamefont {A.}~\bibnamefont
  {Salandrino}}, \ and\ \bibinfo {author} {\bibfnamefont {N.}~\bibnamefont
  {Engheta}},\ }\href@noop {} {\bibfield  {journal} {\bibinfo  {journal}
  {Physical review B}\ }\textbf {\bibinfo {volume} {75}},\ \bibinfo {pages}
  {155410} (\bibinfo {year} {2007})}\BibitemShut {NoStop}%
\bibitem [{\citenamefont {Pendry}\ \emph {et~al.}(1996)\citenamefont {Pendry},
  \citenamefont {Holden}, \citenamefont {Stewart},\ and\ \citenamefont
  {Youngs}}]{Pendry1996}%
  \BibitemOpen
  \bibfield  {author} {\bibinfo {author} {\bibfnamefont {J.~B.}\ \bibnamefont
  {Pendry}}, \bibinfo {author} {\bibfnamefont {A.~J.}\ \bibnamefont {Holden}},
  \bibinfo {author} {\bibfnamefont {W.~J.}\ \bibnamefont {Stewart}}, \ and\
  \bibinfo {author} {\bibfnamefont {I.}~\bibnamefont {Youngs}},\ }\href
  {\doibase 10.1103/PhysRevLett.76.4773} {\bibfield  {journal} {\bibinfo
  {journal} {Phys. Rev. Lett.}\ }\textbf {\bibinfo {volume} {76}},\ \bibinfo
  {pages} {4773} (\bibinfo {year} {1996})}\BibitemShut {NoStop}%
\bibitem [{\citenamefont {Pendry}\ \emph {et~al.}(1999)\citenamefont {Pendry},
  \citenamefont {Holden}, \citenamefont {Robbins},\ and\ \citenamefont
  {Stewart}}]{Pendry1999}%
  \BibitemOpen
  \bibfield  {author} {\bibinfo {author} {\bibfnamefont {J.~B.}\ \bibnamefont
  {Pendry}}, \bibinfo {author} {\bibfnamefont {A.~J.}\ \bibnamefont {Holden}},
  \bibinfo {author} {\bibfnamefont {D.~J.}\ \bibnamefont {Robbins}}, \ and\
  \bibinfo {author} {\bibfnamefont {W.}~\bibnamefont {Stewart}},\ }\href@noop
  {} {\bibfield  {journal} {\bibinfo  {journal} {IEEE transactions on microwave
  theory and techniques}\ }\textbf {\bibinfo {volume} {47}},\ \bibinfo {pages}
  {2075} (\bibinfo {year} {1999})}\BibitemShut {NoStop}%
\bibitem [{\citenamefont {Kaina}\ \emph {et~al.}(2015)\citenamefont {Kaina},
  \citenamefont {Lemoult}, \citenamefont {Fink},\ and\ \citenamefont
  {Lerosey}}]{Kaina2015}%
  \BibitemOpen
  \bibfield  {author} {\bibinfo {author} {\bibfnamefont {N.}~\bibnamefont
  {Kaina}}, \bibinfo {author} {\bibfnamefont {F.}~\bibnamefont {Lemoult}},
  \bibinfo {author} {\bibfnamefont {M.}~\bibnamefont {Fink}}, \ and\ \bibinfo
  {author} {\bibfnamefont {G.}~\bibnamefont {Lerosey}},\ }\href@noop {}
  {\bibfield  {journal} {\bibinfo  {journal} {Nature}\ }\textbf {\bibinfo
  {volume} {525}},\ \bibinfo {pages} {77} (\bibinfo {year} {2015})}\BibitemShut
  {NoStop}%
\bibitem [{\citenamefont {Lanoy}\ \emph {et~al.}(2017)\citenamefont {Lanoy},
  \citenamefont {Page}, \citenamefont {Lerosey}, \citenamefont {Lemoult},
  \citenamefont {Tourin},\ and\ \citenamefont {Leroy}}]{Lanoy2017}%
  \BibitemOpen
  \bibfield  {author} {\bibinfo {author} {\bibfnamefont {M.}~\bibnamefont
  {Lanoy}}, \bibinfo {author} {\bibfnamefont {J.~H.}\ \bibnamefont {Page}},
  \bibinfo {author} {\bibfnamefont {G.}~\bibnamefont {Lerosey}}, \bibinfo
  {author} {\bibfnamefont {F.}~\bibnamefont {Lemoult}}, \bibinfo {author}
  {\bibfnamefont {A.}~\bibnamefont {Tourin}}, \ and\ \bibinfo {author}
  {\bibfnamefont {V.}~\bibnamefont {Leroy}},\ }\href {\doibase
  10.1103/PhysRevB.96.220201} {\bibfield  {journal} {\bibinfo  {journal} {Phys.
  Rev. B}\ }\textbf {\bibinfo {volume} {96}},\ \bibinfo {pages} {220201}
  (\bibinfo {year} {2017})}\BibitemShut {NoStop}%
\bibitem [{\citenamefont {Simovski}\ \emph {et~al.}(2012)\citenamefont
  {Simovski}, \citenamefont {Belov}, \citenamefont {Atrashchenko},\ and\
  \citenamefont {Kivshar}}]{Simovski2012}%
  \BibitemOpen
  \bibfield  {author} {\bibinfo {author} {\bibfnamefont {C.~R.}\ \bibnamefont
  {Simovski}}, \bibinfo {author} {\bibfnamefont {P.~A.}\ \bibnamefont {Belov}},
  \bibinfo {author} {\bibfnamefont {A.~V.}\ \bibnamefont {Atrashchenko}}, \
  and\ \bibinfo {author} {\bibfnamefont {Y.~S.}\ \bibnamefont {Kivshar}},\
  }\href@noop {} {\bibfield  {journal} {\bibinfo  {journal} {Advanced
  Materials}\ }\textbf {\bibinfo {volume} {24}},\ \bibinfo {pages} {4229}
  (\bibinfo {year} {2012})}\BibitemShut {NoStop}%
\bibitem [{\citenamefont {Kaina}\ \emph {et~al.}(2017)\citenamefont {Kaina},
  \citenamefont {Causier}, \citenamefont {Bourlier}, \citenamefont {Fink},
  \citenamefont {Berthelot},\ and\ \citenamefont {Lerosey}}]{Kaina2017}%
  \BibitemOpen
  \bibfield  {author} {\bibinfo {author} {\bibfnamefont {N.}~\bibnamefont
  {Kaina}}, \bibinfo {author} {\bibfnamefont {A.}~\bibnamefont {Causier}},
  \bibinfo {author} {\bibfnamefont {Y.}~\bibnamefont {Bourlier}}, \bibinfo
  {author} {\bibfnamefont {M.}~\bibnamefont {Fink}}, \bibinfo {author}
  {\bibfnamefont {T.}~\bibnamefont {Berthelot}}, \ and\ \bibinfo {author}
  {\bibfnamefont {G.}~\bibnamefont {Lerosey}},\ }\href@noop {} {\bibfield
  {journal} {\bibinfo  {journal} {Scientific reports}\ }\textbf {\bibinfo
  {volume} {7}},\ \bibinfo {pages} {15105} (\bibinfo {year}
  {2017})}\BibitemShut {NoStop}%
\bibitem [{\citenamefont {Lemoult}\ \emph {et~al.}(2010)\citenamefont
  {Lemoult}, \citenamefont {Lerosey}, \citenamefont {de~Rosny},\ and\
  \citenamefont {Fink}}]{Lemoult2010}%
  \BibitemOpen
  \bibfield  {author} {\bibinfo {author} {\bibfnamefont {F.}~\bibnamefont
  {Lemoult}}, \bibinfo {author} {\bibfnamefont {G.}~\bibnamefont {Lerosey}},
  \bibinfo {author} {\bibfnamefont {J.}~\bibnamefont {de~Rosny}}, \ and\
  \bibinfo {author} {\bibfnamefont {M.}~\bibnamefont {Fink}},\ }\href {\doibase
  10.1103/PhysRevLett.104.203901} {\bibfield  {journal} {\bibinfo  {journal}
  {Phys. Rev. Lett.}\ }\textbf {\bibinfo {volume} {104}},\ \bibinfo {pages}
  {203901} (\bibinfo {year} {2010})}\BibitemShut {NoStop}%
\bibitem [{\citenamefont {Lemoult}\ \emph {et~al.}(2013)\citenamefont
  {Lemoult}, \citenamefont {Kaina}, \citenamefont {Fink},\ and\ \citenamefont
  {Lerosey}}]{Lemoult2013}%
  \BibitemOpen
  \bibfield  {author} {\bibinfo {author} {\bibfnamefont {F.}~\bibnamefont
  {Lemoult}}, \bibinfo {author} {\bibfnamefont {N.}~\bibnamefont {Kaina}},
  \bibinfo {author} {\bibfnamefont {M.}~\bibnamefont {Fink}}, \ and\ \bibinfo
  {author} {\bibfnamefont {G.}~\bibnamefont {Lerosey}},\ }\href@noop {}
  {\bibfield  {journal} {\bibinfo  {journal} {Nature Physics}\ }\textbf
  {\bibinfo {volume} {9}},\ \bibinfo {pages} {55} (\bibinfo {year}
  {2013})}\BibitemShut {NoStop}%
\bibitem [{\citenamefont {Lemoult}\ \emph {et~al.}(2011)\citenamefont
  {Lemoult}, \citenamefont {Fink},\ and\ \citenamefont
  {Lerosey}}]{Lemoult2011a}%
  \BibitemOpen
  \bibfield  {author} {\bibinfo {author} {\bibfnamefont {F.}~\bibnamefont
  {Lemoult}}, \bibinfo {author} {\bibfnamefont {M.}~\bibnamefont {Fink}}, \
  and\ \bibinfo {author} {\bibfnamefont {G.}~\bibnamefont {Lerosey}},\ }\href
  {\doibase 10.1080/17455030.2011.611836} {\bibfield  {journal} {\bibinfo
  {journal} {Waves in Random and Complex Media}\ }\textbf {\bibinfo {volume}
  {21}},\ \bibinfo {pages} {591} (\bibinfo {year} {2011})}\BibitemShut
  {NoStop}%
\bibitem [{\citenamefont {Belov}\ \emph {et~al.}(2005)\citenamefont {Belov},
  \citenamefont {Simovski},\ and\ \citenamefont {Ikonen}}]{Belov2005}%
  \BibitemOpen
  \bibfield  {author} {\bibinfo {author} {\bibfnamefont {P.~A.}\ \bibnamefont
  {Belov}}, \bibinfo {author} {\bibfnamefont {C.~R.}\ \bibnamefont {Simovski}},
  \ and\ \bibinfo {author} {\bibfnamefont {P.}~\bibnamefont {Ikonen}},\ }\href
  {\doibase 10.1103/PhysRevB.71.193105} {\bibfield  {journal} {\bibinfo
  {journal} {Phys. Rev. B}\ }\textbf {\bibinfo {volume} {71}},\ \bibinfo
  {pages} {193105} (\bibinfo {year} {2005})}\BibitemShut {NoStop}%
\bibitem [{\citenamefont {Geim}\ and\ \citenamefont
  {Novoselov}(2010)}]{Geim2010}%
  \BibitemOpen
  \bibfield  {author} {\bibinfo {author} {\bibfnamefont {A.~K.}\ \bibnamefont
  {Geim}}\ and\ \bibinfo {author} {\bibfnamefont {K.~S.}\ \bibnamefont
  {Novoselov}},\ }in\ \href@noop {} {\emph {\bibinfo {booktitle} {Nanoscience
  and Technology: A Collection of Reviews from Nature Journals}}}\ (\bibinfo
  {publisher} {World Scientific},\ \bibinfo {year} {2010})\ pp.\ \bibinfo
  {pages} {11--19}\BibitemShut {NoStop}%
\bibitem [{\citenamefont {Robillard}\ \emph {et~al.}(2011)\citenamefont
  {Robillard}, \citenamefont {Bucay}, \citenamefont {Deymier}, \citenamefont
  {Shelke}, \citenamefont {Muralidharan}, \citenamefont {Merheb}, \citenamefont
  {Vasseur}, \citenamefont {Sukhovich},\ and\ \citenamefont
  {Page}}]{robillard2011}%
  \BibitemOpen
  \bibfield  {author} {\bibinfo {author} {\bibfnamefont {J.-F.}\ \bibnamefont
  {Robillard}}, \bibinfo {author} {\bibfnamefont {J.}~\bibnamefont {Bucay}},
  \bibinfo {author} {\bibfnamefont {P.}~\bibnamefont {Deymier}}, \bibinfo
  {author} {\bibfnamefont {A.}~\bibnamefont {Shelke}}, \bibinfo {author}
  {\bibfnamefont {K.}~\bibnamefont {Muralidharan}}, \bibinfo {author}
  {\bibfnamefont {B.}~\bibnamefont {Merheb}}, \bibinfo {author} {\bibfnamefont
  {J.}~\bibnamefont {Vasseur}}, \bibinfo {author} {\bibfnamefont
  {A.}~\bibnamefont {Sukhovich}}, \ and\ \bibinfo {author} {\bibfnamefont
  {J.}~\bibnamefont {Page}},\ }\href@noop {} {\bibfield  {journal} {\bibinfo
  {journal} {Physical Review B}\ }\textbf {\bibinfo {volume} {83}},\ \bibinfo
  {pages} {224301} (\bibinfo {year} {2011})}\BibitemShut {NoStop}%
\bibitem [{\citenamefont {Sukhovich}\ \emph {et~al.}(2009)\citenamefont
  {Sukhovich}, \citenamefont {Merheb}, \citenamefont {Muralidharan},
  \citenamefont {Vasseur}, \citenamefont {Pennec}, \citenamefont {Deymier},\
  and\ \citenamefont {Page}}]{sukhovich2009}%
  \BibitemOpen
  \bibfield  {author} {\bibinfo {author} {\bibfnamefont {A.}~\bibnamefont
  {Sukhovich}}, \bibinfo {author} {\bibfnamefont {B.}~\bibnamefont {Merheb}},
  \bibinfo {author} {\bibfnamefont {K.}~\bibnamefont {Muralidharan}}, \bibinfo
  {author} {\bibfnamefont {J.}~\bibnamefont {Vasseur}}, \bibinfo {author}
  {\bibfnamefont {Y.}~\bibnamefont {Pennec}}, \bibinfo {author} {\bibfnamefont
  {P.}~\bibnamefont {Deymier}}, \ and\ \bibinfo {author} {\bibfnamefont
  {J.}~\bibnamefont {Page}},\ }\href@noop {} {\bibfield  {journal} {\bibinfo
  {journal} {Physical review letters}\ }\textbf {\bibinfo {volume} {102}},\
  \bibinfo {pages} {154301} (\bibinfo {year} {2009})}\BibitemShut {NoStop}%
\bibitem [{\citenamefont {Mansfield}\ and\ \citenamefont
  {Kino}(1990)}]{Mansfield1990}%
  \BibitemOpen
  \bibfield  {author} {\bibinfo {author} {\bibfnamefont {S.~M.}\ \bibnamefont
  {Mansfield}}\ and\ \bibinfo {author} {\bibfnamefont {G.~S.}\ \bibnamefont
  {Kino}},\ }\href@noop {} {\bibfield  {journal} {\bibinfo  {journal} {Applied
  Physics Letters}\ }\textbf {\bibinfo {volume} {57}},\ \bibinfo {pages} {2615}
  (\bibinfo {year} {1990})}\BibitemShut {NoStop}%
\bibitem [{\citenamefont {Yves}\ \emph {et~al.}(2017)\citenamefont {Yves},
  \citenamefont {Fleury}, \citenamefont {Berthelot}, \citenamefont {Fink},
  \citenamefont {Lemoult},\ and\ \citenamefont {Lerosey}}]{Yves2017}%
  \BibitemOpen
  \bibfield  {author} {\bibinfo {author} {\bibfnamefont {S.}~\bibnamefont
  {Yves}}, \bibinfo {author} {\bibfnamefont {R.}~\bibnamefont {Fleury}},
  \bibinfo {author} {\bibfnamefont {T.}~\bibnamefont {Berthelot}}, \bibinfo
  {author} {\bibfnamefont {M.}~\bibnamefont {Fink}}, \bibinfo {author}
  {\bibfnamefont {F.}~\bibnamefont {Lemoult}}, \ and\ \bibinfo {author}
  {\bibfnamefont {G.}~\bibnamefont {Lerosey}},\ }\href@noop {} {\bibfield
  {journal} {\bibinfo  {journal} {Nature communications}\ }\textbf {\bibinfo
  {volume} {8}},\ \bibinfo {pages} {16023} (\bibinfo {year}
  {2017})}\BibitemShut {NoStop}%
\end{thebibliography}%

\end{document}